%% file: main.tex
\RequirePackage[OT1]{fontenc}   
\documentclass[conference]{IEEEtran}
\IEEEoverridecommandlockouts
\usepackage{cite}
\usepackage{amsmath,amssymb,amsfonts}
\usepackage{algorithmic}
\usepackage{graphicx}
\usepackage{textcomp}
\usepackage[dvipsnames]{xcolor}

\usepackage[a4paper,left=1.43cm,right=1.43cm,top=1.8cm,bottom=4.43cm]{geometry} 
\usepackage[font=footnotesize]{caption}
\usepackage[font=footnotesize]{subcaption}      
\usepackage{booktabs}
\usepackage{tcolorbox}
\usepackage{bm}			
\usepackage[per-mode=symbol]{siunitx}    		
\usepackage{cite} 
\DeclareSIUnit{\dBm}{dBm}	
\DeclareSIUnit{\eq}{eq}	    
\DeclareSIUnit{\sqrtW}{\ensuremath{\sqrt{\text{W}}}}
\usepackage{layouts}
\usepackage{overpic}		
\definecolor{RDlightgreen}{RGB}{141 192 69}
\definecolor{RDgreen}{RGB}{93 109 68}
\definecolor{mygray}{gray}{0.6}
\usepackage[    disable,            
                backgroundcolor=RDlightgreen!70,
                textcolor=black,
                bordercolor=RDgreen,
                textsize=tiny,]{todonotes}
\usepackage{enumerate}  
\newcommand{\poisson}{{\rm Poisson}}

\usepackage[xindy]{glossaries}
\makenoidxglossaries%
\loadglsentries{abbr}

\usepackage{balance}
\usepackage{flushend}

\tcbset{shield externalize} 

  \usepackage{pgfplots}
  \pgfplotsset{compat=newest}
    \usepackage{shellesc}
  
\usepackage{ifthen}
\newcommand{\externalizeFigures}{false}
\ifthenelse{\equal{\externalizeFigures}{true}}
{
  \usepgfplotslibrary{external}
  \tikzexternalize[prefix=externalized/]
}
{}

\newcommand{\mrm}[1]{ \mathrm{#1} }

\newcommand{\transp}{\mathsf{T}}
\DeclareMathOperator{\diag}{diag}
\DeclareMathOperator{\lognormal}{lognormal}

\newcommand{\vm}[1]{\ensuremath{\bm{#1}}}

\newcommand{\changesBD}[1]{#1}    

\newcommand{\numberScatterers}{\ensuremath{M_\mathrm{sc}}} 
\newcommand{\rcs}{\ensuremath{\sigma}} 

  \usepackage{pgfplots}
  \pgfplotsset{compat=newest}
  \usetikzlibrary{plotmarks}
  \usetikzlibrary{arrows.meta}
  \usepgfplotslibrary{patchplots}
  \usepackage{grffile}
  \usepackage{amsmath}
  \usepackage{tikzscale}		
  
  \newlength{\plotWidth}		

\def\BibTeX{{\rm B\kern-.05em{\sc i\kern-.025em b}\kern-.08em
    T\kern-.1667em\lower.7ex\hbox{E}\kern-.125emX}}
\begin{document}

\title{Location-based Initial Access for Wireless Power Transfer with Physically Large Arrays}

\author{\IEEEauthorblockN{
Benjamin J. B. Deutschmann\IEEEauthorrefmark{4},   
Thomas Wilding\IEEEauthorrefmark{4},   
Erik G. Larsson\IEEEauthorrefmark{2},    
Klaus Witrisal\IEEEauthorrefmark{4}     
}                                     
\thanks{The project has received funding from the European Union’s Horizon 2020 research and
innovation programme under grant agreement No 101013425.}
\IEEEauthorblockA{\IEEEauthorrefmark{4}
Graz University of Technology, Austria}
\IEEEauthorblockA{\IEEEauthorrefmark{2}
Link\"oping University, Sweden}
\IEEEauthorblockA{ \emph{benjamin.deutschmann@tugraz.at} }
}

\maketitle

\begin{abstract}
\Gls{rf} \gls{wpt} is a promising technology for 6G use cases. 
It enables a massive yet sustainable deployment of batteryless \gls{en} devices at an unprecedented scale.
Recent research on 6G is exploring high operating frequencies up to the THz spectrum, where antenna arrays with large apertures are capable of forming narrow, ``laser-like'' beams.
At sub-10\,GHz frequencies, physically large antenna arrays are considered that are operating in the array near field. 
Transmitting spherical wavefronts, power can be focused to a focal point rather than a beam, which allows for efficient and radiation-safe WPT.
We formulate a multipath channel model comprising specular components and diffuse scattering to find the WPT power budget in a realistic indoor scenario.
Specular components can be predicted by means of a geometric model. This is used to transmit power via multiple beams simultaneously, increasing the available power budget and  expanding the initial access distance.
We show that  exploiting this ``beam diversity'' reduces the required fading margin for the initial access to EN devices.
\end{abstract}


\section{Introduction}

\glsresetall            

\changesBD{Current research on 6G is exploring a wide range of new solutions comprising both massive distributed \gls{mimo} systems for sub-10\,\SI{}{\giga\hertz}, and the use of operating frequencies reaching up to the mmWave and sub-THz bands.}
\Gls{rf} \gls{wpt} with antenna arrays at such high frequencies allows forming narrow beams~\cite{Zhang2021} and, due to their large Fraunhofer distance, operating in the array near field.
These electrically large arrays enable focusing power in a point rather than a beam. 
The H2020 project REINDEER\todo{formulation? tw: We envision.../and are envisioned...?} 
envisions the use of physically large, and distributed antenna arrays for indoor use cases~\cite{D1_1}. 
This wireless infrastructure, termed \gls{rw}, will inherently support the use of \gls{en} devices, which are batteryless and solely powered through WPT.
\changesBD{That enables a massive, yet sustainable deployment of devices. 
Eventually, RW may be an enabling technology for future generations of \glspl{wsn} and \gls{iot} use cases.}
Since the infrastructure operates at sub-10\,\SI{}{\giga\hertz} frequencies, electrically large apertures are also physically large, which holds some beneficial advantages for WPT:
\begin{enumerate}[i)]
    \item Focusing power in the array near-field can lead to a high power density within the focal point and a low power density outside, even around the transmitting antennas.
    Physically small concentrated arrays typically exhibit higher power densities close to the array.
    \item For  distributed architectures, power is focused within a sphere with a diameter of $\lambda / 2$~\cite{VanderPerre2019}, $\lambda$ being the wavelength at the operating frequency $f$.
    At lower frequencies and thus larger wavelengths, the size of the focal point, as well as the apertures of receiving antennas, are large. 
    \item 
    The maximum power density, limited by human exposure restrictions~\cite{1999-519-EC}, is typically \SI{10}{\watt\per\meter\squared}.
    Large focal points and larger receive antennas thus yield a higher power receivable by EN devices at lower frequencies.
\end{enumerate}

For focusing power on EN devices with downlink precoding schemes like \gls{mrt}, \gls{csi} must be available. 
Channel state estimation can be performed on the first
signal sent by an EN device, possibly through backscatter communication~\cite{Kashyap2015a}. 
However, in the initial access phase, the EN device has to be supplied with sufficient power to exceed the device sensitivity, i.e., the minimum power required for wake-up and backscatter communication~\cite{D2_1}.
One possible approach for initial access is beam sweeping, where the transmit array sweeps beams sequentially according to a predefined codebook~\cite{Wu2021} to power up the EN device for the first time. 
Choosing the single beam to achieve the best receive power or \gls{snr} is also known as exploiting the beam selection diversity~\cite{Casas2003}. 
Due to the narrow beamwidth at mmWave bands, some authors propose using dual-band approaches that exploit the link budget advantages and reduced search space of sub-6\,\SI{}{\giga\hertz} frequencies, while using the large bandwidth at mmWave frequencies for high-rate communications~\cite{Polese2017,Giordani2019}. 
Others use radar measurements to aid the beam sweeping with the most likely directions of receiver positions~\cite{Gonzalez-Prelcic2016}. 
In indoor scenarios, environment-awareness could aid the choice for predefined codebooks, assuming that the possible locations of EN devices, as well as the propagation environment, are at least partially known. 
However, beam sweeping in indoor scenarios suffers from fading due to severe multipath propagation, possibly originating from unknown 
objects in the environment.

\gls{rw} supports coherent transmission via multiple beams simultaneously. 
Environment-awareness allows exploiting specular reflections at walls to increase the power budget at the EN device over what is achievable if only a single beam, e.g., only the \gls{los}, was used. 
The main contribution in this paper is the 
demonstration of exploiting a novel form of beam diversity:
in a simultaneous multi-beam transmission from one transmit array (also termed RadioWeaves panel), 
the phases of the individual beams can be varied to reduce the necessary fading margin for the initial access to EN devices.
Environment-awareness and beam diversity enable to make some array gain usable without \gls{csi}, and thus aid to increase the initial access distance of EN devices in indoor environments.
We formulate a multipath channel model to evaluate the WPT power budget and demonstrate the achievable gains through exploiting beam diversity.

The rest of the paper is structured as follows. Section~\ref{sec:channel} describes a geometry based stochastic channel model for WPT. Section~\ref{sec:wpt} investigates WPT schemes by means of numeric simulations. Section~\ref{sec:conclusion} concludes the paper.

\section{Channel Model}\label{sec:channel}
This section presents a channel model that allows a physically accurate representation of the power received by an EN device in a dense multipath  environment.
Typical indoor radio environments envisioned for RW deployments can be described through a channel that exhibits both deterministic, \glspl{smc} and stochastic, diffuse/\gls{dm}~\cite{Wilding2018_AccuracyBounds}.
Operating 
at sub-10\,\SI{}{\giga\hertz} frequencies, the nominal channel bandwidth is typically \SI{20}{\mega\hertz}~\cite{EN_300_328,EN_301_893}. 
Thus, we establish a channel model based on the assumption of a narrowband frequency-flat fading channel used for \gls{mimo} systems~\cite{paulraj2006,molisch2010wireless}.

We consider a single array centered at a position 
$\bm{p}_\mrm{RW}=[x_\mrm{RW},y_\mrm{RW},z_\mrm{RW}]^\mathsf{T}$ 
consisting of $L$ antennas and transmitting power wirelessly\todo{should we keep it more general here? WPT as application} 
to a single EN device at position 
$\bm{p}_\mrm{EN}= [x_\mrm{EN},y_\mrm{EN},z_\mrm{EN}]^\mathsf{T}$. 
The $\ell$\textsuperscript{th} array element position is denoted as  
$\vm{p}^{(\ell)}_{\mrm{RW}}=[x^{(\ell)}_{\mrm{RW}},y^{(\ell)}_{\mrm{RW}},z^{(\ell)}_{\mrm{RW}}]^\mathsf{T}$. 
We employ a memoryless \gls{miso} channel model where the EN device receives a complex 
baseband amplitude, i.e., a phasor, \todo{bd: could be referenced with~cite{Franks1981}} 
\begin{align}\label{eq:channel_model}
    y &= \sum_{k=1}^K \vm{h}^\transp_k \vm{s} + 
    \sum_{k=1}^K \vm{h}^\transp_{\mathrm{sc},k} \vm{s}  
    + {n} 
\end{align}
where $\vm{h}_k = [h_{k,1},\dots,h_{k,L}]^\mathsf{T} \in \mathbb{C}^{L \times 1}$ is the channel vector of the $k$\textsuperscript{th} SMC, and $\vm{s} = [s_{1},\dots,s_{L}]^\mathsf{T} \in \mathbb{C}^{L \times 1}$ the transmit signal vector of the array in complex baseband. 
 
The first term in \eqref{eq:channel_model} models deterministic reflections as a sum of scalar products of the separate channel-vectors for $K$ SMCs, including the LoS.
The second term models \gls{dm} by means of 
point source scatterers where $\vm{h}_{\mathrm{sc},k} \in \mathbb{C}^{L \times 1}$ denotes the scatter channel vector of the $k$\textsuperscript{th} SMC.
The third term $n\in\mathbb{C}$ denotes complex \gls{awgn} with variance $\sigma_\mathrm{n}^2$.
The following sections describe the models for \glspl{smc} and \gls{dm} in detail.

\subsection{Specular Multipath}\label{sec:deterministic}
Deterministic \glspl{smc} are modeled according to an image source model~\cite{Leitinger2015} in combination with an environment floor plan, allowing to compute the position $\vm{p}_{\mrm{RW},k}$ of the image source representing the $k$\textsuperscript{th} \gls{smc}. Note that one of the image sources represents the true RW position $\vm{p}_\mrm{RW}$. 
The $\ell$\textsuperscript{th} element of the channel vector $\vm{h}_k$ is thus modeled as
\begin{align}\label{eq:channel_vector}
    [\bm{h}_{k}]_{\ell} &= 
    \frac{\lambda}{\sqrt{4\pi}} \,
    \frac{1}{\sqrt{4\pi} d_{k,\ell}} \,
    g_{\mrm{SMC},k} \,
    e^{-j \frac{2\pi}{\lambda} d_{k,\ell}} 
\end{align} 
where $d_{k,\ell} = \|\vm{p}_\mrm{EN}-\vm{p}^{(\ell)}_{\mrm{RW},k}\|$ is the distance between the transmit antenna $\ell \in \left\{1,\dots,L\right\}$ of the image source $k$ and the EN device, the operator $\|\cdot\|$ denotes the 
vector norm and $[\,\cdot\,]_{\ell}$ denotes the $\ell$\textsuperscript{th} vector element.
The first factor in \eqref{eq:channel_vector} models the square root of the receiving aperture $A = \frac{\lambda^2}{4\pi}$ assuming a unity gain antenna. 
The second factor models the spread 
of power over the surface of a sphere with a radius $d_{k,\ell}$. 
The factor $g_{\mrm{SMC},k}\in\mathbb{C}$ represents an amplitude gain and phase-shift associated with reflection $k$, e.g., a specific wall, while the exponential term models the phase shift due to the propagation distance $d_{k,\ell}$~\cite{balanis}. 

For simplicity, we assume lossless, isotropic antennas, which exhibit a gain of $G_\ell = 1$. 
Furthermore, we assume no polarization losses and perfect antenna impedance matching of both the transmit and receive antennas. 
Under these assumptions, 
the channel vector from \eqref{eq:channel_vector} effectively models the Friis transmission equation for power wave amplitudes. 
That is, if the transmit signals $[\bm{s}]_\ell$ are power waves of the dimension $\sqrt{\text{power}}$, 
the power received by the EN device is~\cite{Carlin56}
\begin{align}\label{eq:receive_power}
    P_\mrm{RX} = \left|y \right|^2 .
\end{align}
The entries $[\bm{h}_k]_\ell$ of the channel vector are dimensionless transmission coefficients, i.e., S-parameters.

\begin{figure}[tb!]	
	\centering
	\setlength{\plotWidth}{1.05\linewidth}
	    \ifdefined\ExcludeFigures   
            \ExcludeFigures
        \else
            \input{figures/setupICC.tex} 
        \fi 
    	\caption{Simulation scenario with four walls, as well as a floor. 
	A $(\SI{5}{\meter} \times \SI{9}{\meter}\times \SI{3.5}{\meter})$ large room is simulated. 
	Virtual mirror arrays \changesBD{with their local coordinate systems} are indicated in the figure along with the EN device. Scatter points are distributed along the front wall in the direction of LoS to the EN device.}
	\label{fig:setup}
\end{figure}
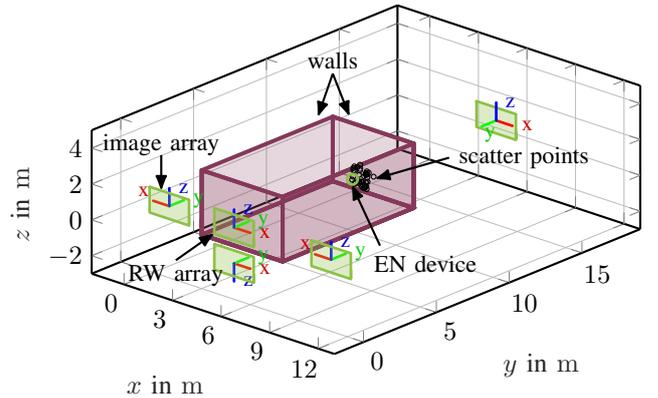	

\subsection{Diffuse Multipath}\label{sec:stochastic}
The second term from \eqref{eq:channel_model} 
represents stochastic scattering at small, distributed objects, or surfaces that are rough with respect to the signal wavelength $\lambda$~\cite{KulmerPIMRC2018}.
The \gls{dm} is modeled by a random number of $\numberScatterers$ point scatterers at positions $\vm{p}_{\mrm{sc},m} = [x^{(m)}_\mrm{sc},y^{(m)}_\mrm{sc},z^{(m)}_\mrm{sc}]^\mathsf{T}$ where all impinging specular waves are rescattered, assuming only single-bounce scattering. 
The resulting channel vectors for each \gls{smc} $k$ are defined as
\begin{align}\label{eq:channel_vector_sc}
    \bm{h}^{\transp}_{\mathrm{sc},k} &= 
        \bm{h}^{\transp}_\mrm{RX} \,
        \bm{\Sigma}_\mrm{sc} \,
        \bm{H}_{\mrm{TX},k}
\end{align}
where $\bm{H}_{\mrm{TX},k}\in\mathbb{C}^{\numberScatterers \times L}$ is the channel matrix from the $k$\textsuperscript{th} mirror array to the scatterers and $\bm{h}_\mrm{RX}\in\mathbb{C}^{\numberScatterers\times 1}$ is the channel vector from the scatterers to the EN device. 
The matrix 
$\bm{\Sigma}_\mrm{sc} = \diag\left\{ [\sqrt{\rcs_{\scalebox{.6}{1}}}e^{j\varphi_1}, \dots,\sqrt{\rcs_{\scalebox{.6}{\numberScatterers}}}e^{j\varphi_{\numberScatterers}}]\right\}\in\mathbb{C}^{\numberScatterers \times \numberScatterers}$ is a diagonal matrix containing the 
\gls{rcs} $\rcs_{m}$ of point source scatterer $m \in \left\{1,\dots,\numberScatterers\right\}$ and a respective phase-shift $\varphi_{m}$ on its main diagonal.
The \glspl{rcs} are modeled i.i.d.\todo{tw: check if good notation} 
log-normally~\cite{Santos2010,Gustafson2014}, i.e., $\rcs_{m}~\sim~\lognormal(\breve{\mu}_\mrm{sc},\breve{\sigma}_\mrm{sc}^{2})$, with 
$\breve{\mu}_\mrm{sc}=\ln\Big(\frac{\mu_\mrm{sc}^2}{\sqrt{\sigma_\mrm{sc}^2+\mu_\mrm{sc}^2}}\Big)$ and 
$\breve{\sigma}_\mrm{sc}^2=\ln\left(\frac{\sigma_\mrm{sc}^2}{\mu_\mrm{sc}^2}+1\right)$.
The phase-shifts $\varphi_m$ are i.i.d. uniform, i.e., $\varphi_{m}~\sim~\mathcal{U}(0,\,2\pi)$. \todo{can $\sim$ be used for a realization instead of a distribution?}
Let $d_{k,\ell,m} = \|\vm{p}_{\mrm{sc},m}-\vm{p}_{\mrm{RW},k}^{(\ell)}\|$ be the distance between scatterer $m$ and transmit antenna $\ell$ of the $k$\textsuperscript{th} mirror array, the $(\ell,m)$\textsuperscript{th} entry of the channel matrix $\bm{H}_{\mathrm{TX},k}$ is given as 
\begin{align}\label{eq:TX_channel_matrix}
    [\bm{H}_{\mathrm{TX},k}]_{\ell,m} = 
    \frac{1}{\sqrt{4\pi}\, d_{k,\ell,m}} \,
    g_{\mrm{SMC},k} \,
    e^{-j \frac{2\pi}{\lambda} d_{k,\ell,m}}.
\end{align}
Consequently, the squared product of the RW channel matrix and the transmit power wave, i.e., 
$\left| [\bm{H}_{\mathrm{TX},k} \bm{s}]_m\right|^2$, is the power density at the location of the scatterer $m$ caused by the signal transmitted from image array $k$. 
This power density is rescattered isotropically (with respect to the receiver~\cite{balanis}) by the scatterer with a radar cross-section $\rcs_{m}$.
The scatterer to EN device channel is represented by the vector $\bm{h}_\mrm{RX}$, which models the gain of the power wave propagating from scatterer $m$ to the EN device as
\begin{align}\label{eq:RX_channel_matrix}
    [\bm{h}_\mrm{RX}]_{m} = 
    \frac{\lambda}{\sqrt{4\pi}} \,
    \frac{1}{\sqrt{4\pi} d_{m}} \,
    e^{-j \frac{2\pi}{\lambda} d_{m}}
\end{align}
where $d_{m} = \|\vm{p}_{\mrm{sc},m}-\vm{p}\|$ is the distance between the EN device and scatterer $m$.

\begin{table}[tb!]
\caption{List of Simulation Parameters}
\begin{center}
\begin{tabular}{ l c c c } \toprule[0.75pt] 
\textbf{Variable} & \textbf{Symbol} & \textbf{Unit} & \textbf{Value} \\[0.75pt] \hline \addlinespace[2pt] 
Carrier frequency & $f$ & \SI{}{\giga\hertz} & \SI{2.4}{}   \\
URA width & $l_x$ & \SI{}{\meter} & \SI{2.5}{}   \\
URA height & $l_z$ & \SI{}{\meter} & \SI{1.5}{}   \\
Center pos. URA & $\bm{p}_\mrm{RW}$ & \SI{}{\meter} & $\left[\SI{5}{}, \SI{0}{}, \SI{1}{}\right]^\transp$   \\
Pos. EN device & $\bm{p}_\mrm{EN}$ & \SI{}{\meter} & $\left[\SI{5}{}, \SI{8.125}{}, \SI{1}{}\right]^\transp$   \\
Center pos. ellipsoid & $\bm{p}_\mrm{sc}$ & \SI{}{\meter} & $\left[\SI{5}{}, \SI{8.75}{}, \SI{1}{}\right]^\transp$   \\
Number of transmit antennas & $L$ & - & \SI{960}{}  \\
Poisson mean & $\lambda_\mrm{sc}$ & \SI{}{\meter^{-3}} & \SI{10}{} \\ 
Number of scatter points & $\numberScatterers$ & - & \SI{38}{}  \\
SMC amplitude gain & $|g_{\mrm{SMC},k}|$ & \SI{}{\dB} & \SI{-3}{}   \\
Lognormal mean & $\mu_\mrm{sc}$ & \SI{}{\centi\meter\squared} & $\SI{10}{}^2\pi$   \\
Lognormal std. dev. & $\sigma_\mrm{sc}$ & \SI{}{\centi\meter\squared} & $\SI{20}{}\pi$   \\
\bottomrule[0.75pt]
\end{tabular}
\label{tab:sim_parameters}
\end{center}
\end{table}

The scatter channel vector described by~(\ref{eq:channel_vector_sc}) thus effectively models the bistatic radar range equation for power wave amplitudes. 
Note that  the product $\bm{\Sigma}_\mrm{sc} \bm{H}_{\mrm{TX},k} $ is dimensionless and correct in terms of power, and thus a matrix of S-parameters.
The receive channel vector $\bm{h}_\mrm{RX}$ is dimensionless and an S-parameter vector by itself.
Hence, under the assumptions made, the model in~(\ref{eq:channel_model}) 
is physically correct in terms of power and thus allows to compute the power budget at an EN device location for a given scenario. 
The scatter channel vector attributable to each point source scatterer can be seen as a pin-hole channel~\cite{paulraj2006}.

\section{Wireless Power Transfer}\label{sec:wpt}

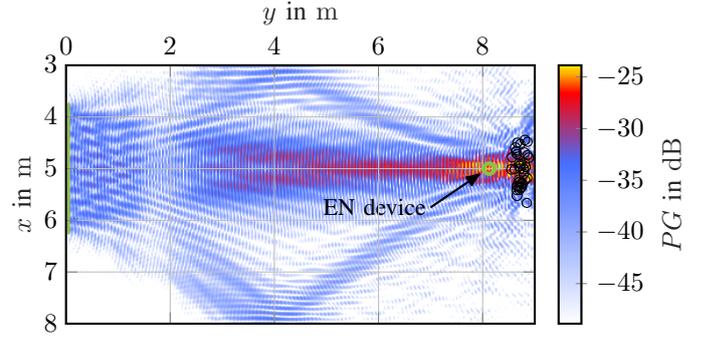
\begin{figure}[bt!]	
    	\centering
         \setlength{\plotWidth}{0.68\textwidth}
	    \ifdefined\ExcludeFigures   
            \ExcludeFigures
        \else
            \input{./figures/MRT_full.tex}
        \fi 

    	\caption{
    	Path gain $PG$ evaluated on a cutting plane (at $z=\SI{1}{\meter}$, perpendicular to the center of the array) through the simulated room. The simulation parameters used are listed in Table~\ref{tab:sim_parameters}. MRT has been used for precoding, assuming perfect CSI including point scatterers.
    	Note that the computed PG very close to the scatter points is incorrect, due to a violation of model assumptions.
    	}
	\label{fig:MRT_full}
\end{figure}	

\begin{figure*}[tb!]	
     \hspace{1mm}
     \begin{subfigure}[t]{0.29\textwidth}
        \vskip 0pt	
         \centering
         \setlength{\plotWidth}{0.8\textwidth}
	    \ifdefined\ExcludeFigures   
	                \ExcludeFigures
        \else
            \includegraphics{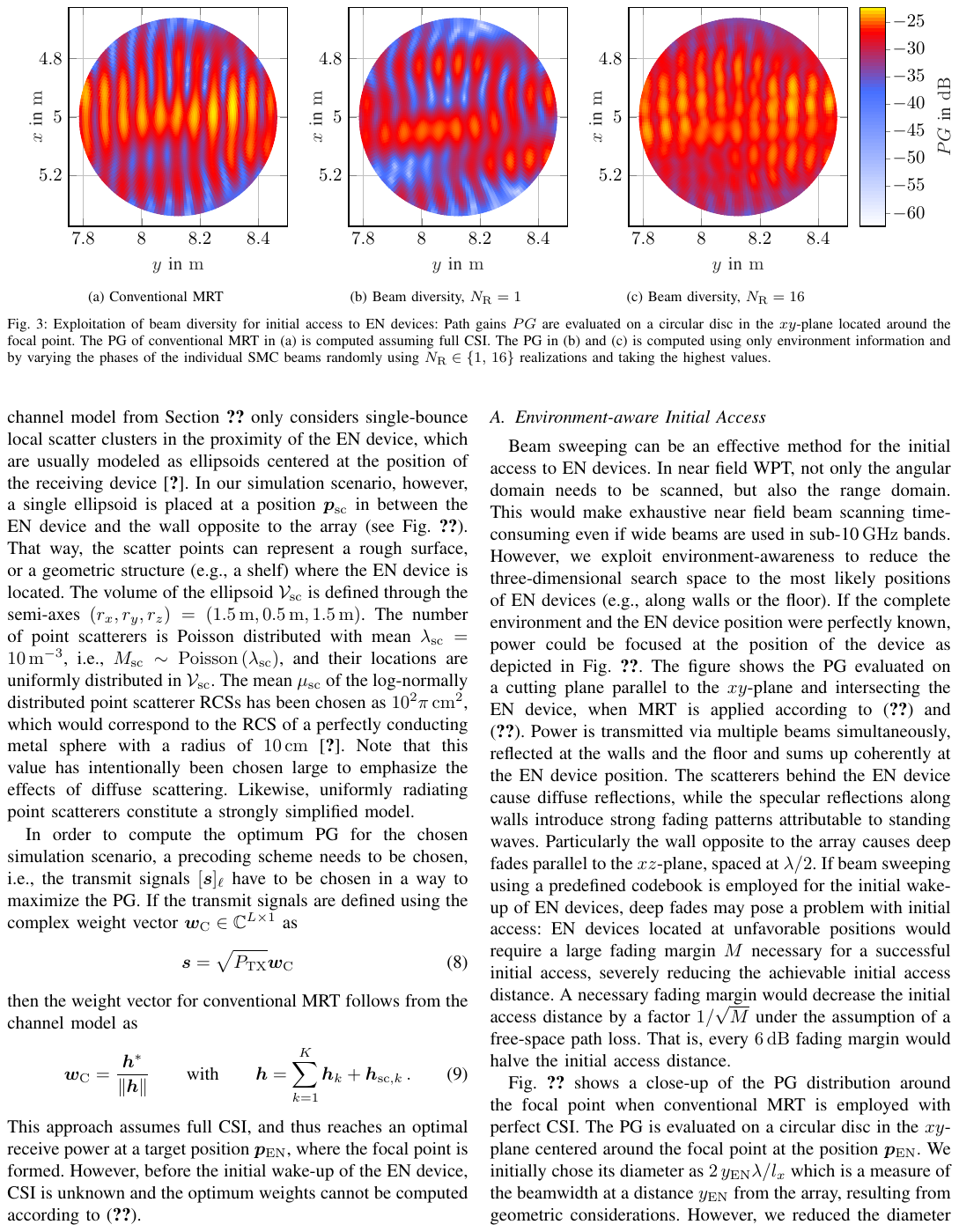}
        \fi 
         \caption{Conventional MRT}
         \label{fig:MRT_domain}
     \end{subfigure}
     \begin{subfigure}[t]{0.29\textwidth}
        \vskip 0pt	
         \centering
         \setlength{\plotWidth}{0.8\textwidth}	
	    \ifdefined\ExcludeFigures   
            \ExcludeFigures
        \else
            \includegraphics{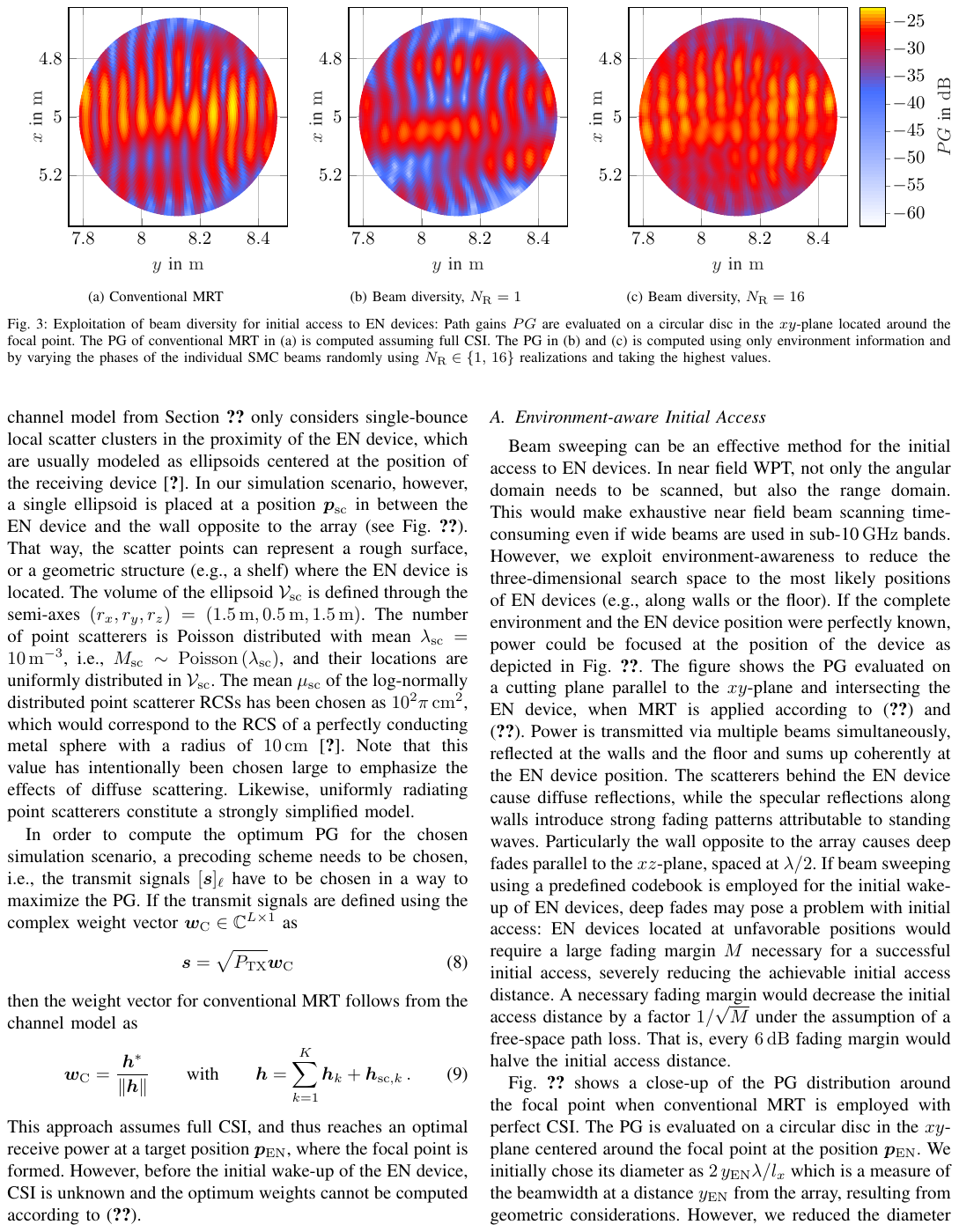}
        \fi 
         	
         \caption{Beam diversity, $N_\mrm{R}=1$}
         \label{fig:M1_domain}
     \end{subfigure}
     \begin{subfigure}[t]{0.29\textwidth}
         \vskip 0pt	
         \centering
                  \setlength{\plotWidth}{0.8\textwidth} 
	    \ifdefined\ExcludeFigures   
            \ExcludeFigures
        \else
             \includegraphics{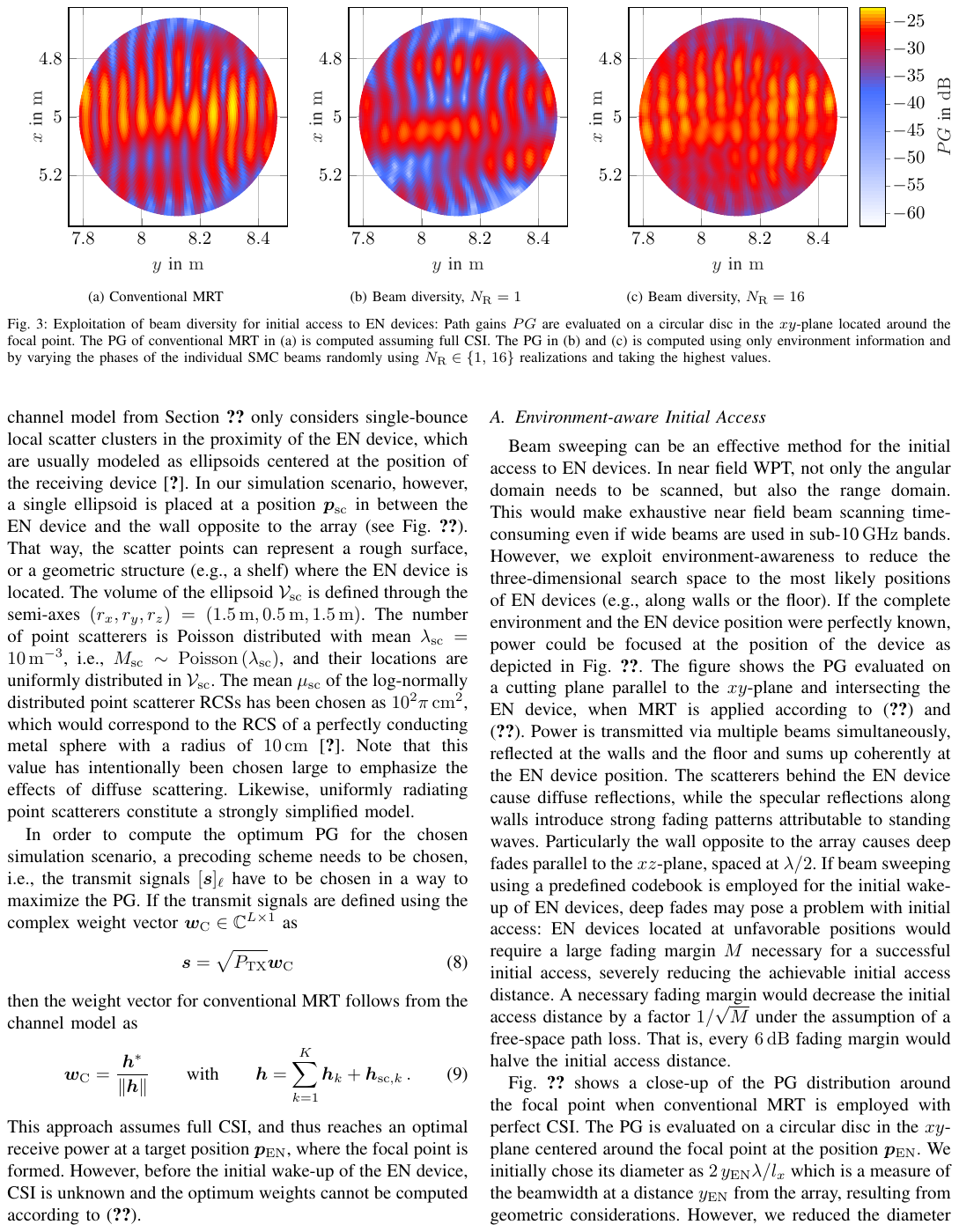}
        \fi 
         \caption{Beam diversity, $N_\mrm{R}=16$}
         \label{fig:M16_domain}
     \end{subfigure}
     \begin{subfigure}[t]{0.058\textwidth}
         \vskip 0pt	
         \centering
                  \setlength{\plotWidth}{1\textwidth}	
         	\input{figures/CB.tex}
     \end{subfigure}
    	\caption{
    	Exploitation of beam diversity for initial access to EN devices: 
    	Path gains $PG$ are evaluated on a circular disc in the $xy$-plane located around the focal point. 
    	The PG of conventional MRT in (a) is computed assuming full CSI.
    	The PG in (b) and (c) is computed using only environment information and by
    	varying the phases of the individual SMC beams randomly using $N_\mrm{R} \in \left\{1,\, 16\right\}$ realizations and taking the highest values.
    	}\label{fig:circular_domains}
\end{figure*}	

In the following, we use the channel model in (\ref{eq:channel_model}) to compute the power budget for the simulation scenario in Fig.~\ref{fig:setup}. 
Particularly, under the assumptions made, we evaluate the \gls{pg} at the position $\bm{p}_\mrm{EN}$ of the EN device as
\begin{align}
    PG = \frac{P_\mrm{RX}}{P_\mrm{TX}} 
\end{align}
where $P_\mrm{TX}$ is the transmit power, i.e., $P_\mrm{TX} = \|\vm{s}\|^2$. 
For all subsequent simulations, we use the parameters listed in Table~\ref{tab:sim_parameters}, some of which are chosen for this theoretical analysis from an educated guess and may vary in an actual deployment scenario. 

Despite the fact that \gls{rw} supports massive, distributed antenna arrays, we use only one physically large \gls{ura} for simplicity, with an inter-element spacing of $\lambda/2$. 
The array is mounted on one side of the simulated room and has the physical dimensions $\left( l_x \times l_z \right) = {(2.5 \times 1.5)}\SI{}{\metre\squared}$. 
It is located at the position $\bm{p}_\mrm{RW}$ inside the room and is mirrored at the walls to obtain image sources (c.f., Section~\ref{sec:deterministic}). 
We only consider first-order image sources in the simulations. 
From the perspective of the EN device, the incident signals are virtually impinging from the positions 
of the actual array and the four corresponding image arrays, each associated with an SMC amplitude gain $|g_{\mrm{SMC},k}|$. 
We have chosen $g_{\mrm{SMC},k}$ to be real-valued and constant $\SI{-3}{\dB}$ for all $K$ walls. 
\changesBD{
This value has been chosen larger than a typical value for reflections at concrete walls (e.g., $\SI{-6}{\dB}$ \cite{Bois2000}) 
to emphasise the effects of local fading.} 
The stochastic channel model from Section~\ref{sec:stochastic} only considers single-bounce local scatter clusters in the proximity of the EN device, 
which are usually modeled as ellipsoids centered at the position of the receiving device~\cite{COST2100}.
In our simulation scenario, however, a single ellipsoid is placed at a position $\bm{p}_\mrm{sc}$ in between the EN device and the wall opposite to the array (see Fig.~\ref{fig:setup}). 
That way, the scatter points can represent a rough surface, or a geometric structure (e.g., a shelf) where the EN device is located.
The volume of the ellipsoid $\mathcal{V}_\mrm{sc}$ is defined through the semi-axes $\left(r_x, r_y, r_z\right) = \left(\SI{1.5}{\meter}, \SI{0.5}{\meter}, \SI{1.5}{\meter}\right)$.
The number of point scatterers is Poisson distributed with mean $\lambda_\mrm{sc} = \SI{10}{\meter^{-3}}$, i.e., 
$\numberScatterers \sim \poisson\left(\lambda_\mrm{sc}\right)$, \todo{here again, we have a sample rather than a distribution..}
and their locations are uniformly distributed in $\mathcal{V}_\mrm{sc}$.
The mean $\mu_\mrm{sc}$ of the log-normally distributed point scatterer RCSs has been chosen as $10^2\pi\,\SI{}{\centi\meter\squared}$, which would correspond to the RCS of a perfectly conducting metal sphere with a radius of $\SI{10}{\centi\meter}$~\cite{kingsley1999}.
Note that this value has intentionally been chosen large to emphasize the effects of diffuse scattering.
Likewise, uniformly radiating point scatterers constitute a strongly simplified model.

In order to compute the optimum 
PG for the chosen simulation scenario, a precoding scheme needs to be chosen, i.e., the transmit signals $[\bm{s}]_\ell$ have to be chosen in a way to maximize the PG.
If the transmit signals are defined using the complex weight vector $\bm{w}_\mrm{C}\in\mathbb{C}^{L\times 1}$ as
\begin{align}\label{eq:s_MRT}
    \bm{s} = \sqrt{P_\mrm{TX}} \bm{w}_\mrm{C} 
\end{align}
then the weight vector for conventional 
MRT follows from the channel model as
\begin{align}\label{eq:C_MRT}
    \bm{w}_\mrm{C} = \frac{\bm{h}^{\ast}}{\|\bm{h}\|}
    \qquad\text{with}\qquad  
    \bm{h} = \sum_{k=1}^K \bm{h}_k + \bm{h}_{\mrm{sc},k} \, .
\end{align}
This approach assumes full CSI, and thus reaches an optimal receive power at 
a target position $\bm{p}_\mrm{EN}$, where the focal point is formed. \todo{bd: Is this formulation okay? "focal point" is to be introduced.}
However, before the initial wake-up of the EN device, CSI is unknown and the optimum weights cannot be computed according to \eqref{eq:C_MRT}.

\subsection{Environment-aware Initial Access}
Beam sweeping can be an effective method for the initial access to EN devices.
In near field WPT, not only the angular domain needs to be scanned, but also the range domain.
This would make \changesBD{exhaustive near field beam scanning time-consuming} even if wide beams are used in sub-10\,\SI{}{\giga\hertz} bands.
However, we exploit environment-awareness to reduce the three-dimensional search space to the most likely positions of EN devices (e.g., along walls or the floor).
If the complete environment and the EN device position were perfectly known, power could be focused at the position of the device as depicted in Fig.~\ref{fig:MRT_full}.
The figure shows the PG evaluated on a cutting plane parallel to the $xy$-plane and intersecting the EN device, when MRT is applied according to \eqref{eq:s_MRT} and \eqref{eq:C_MRT}. 
Power is transmitted via multiple beams simultaneously, reflected at the walls and the floor and sums up coherently at the EN device position. 
The scatterers behind the EN device cause diffuse reflections, while the specular reflections along walls introduce strong fading patterns attributable to standing waves. 
Particularly the wall opposite to the array 
causes deep fades parallel to the $xz$-plane, spaced at $\lambda/2$.
If beam sweeping using a predefined codebook is employed for the initial wake-up of EN devices, deep fades may pose a problem with initial access: 
EN devices located at unfavorable positions would require a large fading margin $M$ necessary for a successful initial access, severely reducing the achievable initial access distance.
A necessary fading margin would decrease the initial access distance by a factor $1/\sqrt{M}$ under the assumption of a free-space path loss. 
That is, every $\SI{6}{\dB}$ fading margin would halve the initial access distance.

Fig.~\ref{fig:MRT_domain} shows a close-up of the PG distribution around the focal point when conventional MRT is employed with perfect CSI. 
The PG is evaluated on a circular disc in the $xy$-plane centered around the focal point at the position $\bm{p}_\mrm{EN}$.
\changesBD{We initially chose its diameter as $2\,y_\mrm{EN}\lambda / l_x$ which is a measure of the beamwidth at a distance $y_\mrm{EN}$ from the array, resulting from geometric considerations.
However, we reduced the diameter by a factor of \SI{0.62}{} to omit a performance decrease due to low powers at the disc edges.}
When awareness of the channel model is used for beam sweeping, the deterministic part of the channel $\bm{h}_k$ can be predicted based on a target position $\vm{p}_\mrm{EN}$, using \eqref{eq:channel_vector}. Unfortunately, this will lead to strong local fading, as illustrated in Fig.~\ref{fig:MRT_domain}, 
and a large fading margin is needed to overcome this local fading.

\subsection{Beam Diversity}
The distribution of power in the region surrounding the focal point can be improved through a suitable precoding scheme. 
Beam diversity can be effectively exploited to even out deep fades and generate a smoother power distribution in proximity of the focal point.
As a simple scheme in a multibeam transmission, the phases $\varphi_k$ of SMC beams can be varied to reduce the necessary fading margin at the cost of lower peak powers.
In this regard, we propose to assign equal power to each of the beams to maximally 
affect the local fading. 
That is, we choose the transmit signal $\bm{s} = \sqrt{{P_\mrm{TX}}} \bm{w}_\mrm{BD}$ with the weights defined as 
\begin{align}
    \bm{w}_\mrm{BD} = \frac{\sum_{k=1}^K \bm{w}_{\mrm{BD},k}}{\| \sum_{k=1}^K \bm{w}_{\mrm{BD},k} \|}
    \quad\text{with}\quad
    \bm{w}_{\mrm{BD},k} = \frac{\bm{h}^\ast_k}{\| \bm{h}_k \|} e^{j\varphi_k} .
\end{align}
The individual weight vectors $\bm{w}_{\mrm{BD},k}$ are computed for some target position $\vm{p}_\mrm{EN}$ using \eqref{eq:channel_vector}. 
Fig.~\ref{fig:M1_domain} shows the PG distribution for $N_\mrm{R}=1$ realization of $K$ beam phases $\varphi_k$ drawn from uniform distributions, i.e. $\varphi_k \sim \mathcal{U}(0,\,2\pi)$.
This approach assumes known environment information in terms of the SMC model, c.f. (\ref{eq:channel_vector}), while the scatterer components $\bm{h}_{\mrm{sc},k}$ are unknown. 

\begin{figure}[tb!]	
    	\centering
	\setlength{\plotWidth}{0.85\linewidth}
	    \ifdefined\ExcludeFigures   
            \ExcludeFigures
        \else
            \input{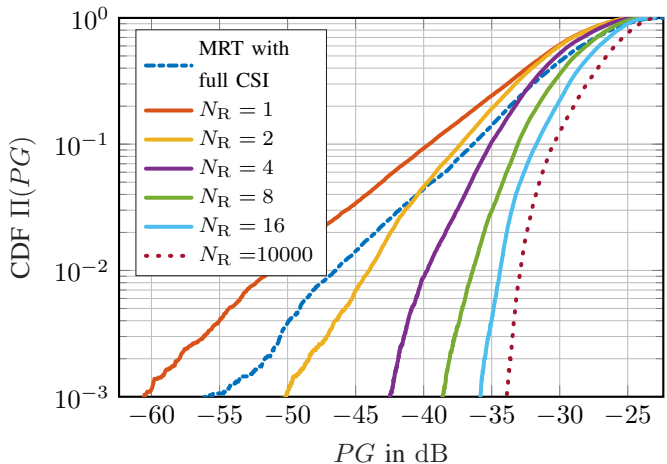}
        \fi 
	
    	\caption{
    	Path gain $PG$ evaluated on the circular disc depicted in Fig.~\ref{fig:circular_domains}.
    	Varying the phases of the individual MPC beams randomly using $N_\mrm{R} \in \{1, 2, 4, 8, 16\}$ realizations and taking the highest values can help to reduce the necessary fading margin for initial access.
    	}\label{fig:CDF}
\end{figure}	
Deep fades still occur, but their depth has been reduced, 
i.e., the use of several realizations of random phases $\varphi_k$ increases the probability of waking up an EN device located at an unfavorable position.
This effect can be further exploited through multiple iterations of the beam diversity scheme: 
Fig.~\ref{fig:M16_domain} shows the maximum PG for every position on the circular disc from $N_\mrm{R}=16$ realizations of $K$ random beam phases.
The smooth power distribution depicted in the figure will not exist at a single time instance, but after $N_\mrm{R}$ attempts, every position has at least once experienced the peak PGs depicted. 
If the power received by an EN device exceeds the device sensitivity for one realization, this is sufficient to transmit a signal on which the \gls{rw} infrastructure can perform channel state estimation.
\changesBD{ 
An exhaustive beam sweep exploiting this scheme may initially take several tens of minutes. 
However, the procedure has to be completed only once, for the initial wake-up of EN devices.}

To analyze the possible gain in initial access distance, we evaluate the \gls{cdf} of the PG distribution across the circular disc. 
We compute the distribution from PGs across an equally-spaced Cartesian grid on the disc.
Fig.~\ref{fig:CDF} depicts the CDFs of the maximum PG of $N_\mrm{R} \in \{1, 2, 4, 8, 16\}$ realizations.
For comparison, the CDF for MRT with perfect CSI (including point scatterers) is depicted, which shows higher peak powers, but also a higher probability of deep fades.
For the first drawn realization ($N_\mrm{R}=1$), the complete range of power values lies below what is achievable with MRT. 
However, drawing more realizations, it is observable that low power values, i.e., deep fades, become less likely.
Targeting an outage probability of less than $\SI{1}{\percent}$, i.e., the horizontal line at $\Pi(PG)=10^{-2}$, the beam diversity scheme using $N_\mrm{R}=16$ realizations would have reduced the necessary fading margin by $\SI{12}{\dB}$ when compared with conventional MRT. That corresponds to an improvement in initial access distance by a factor of $4$, assuming a free-space path loss.
Four repetitions $N_\mrm{R}=4$ reduce the fading margin by $\SI{12}{\dB}$ compared to a single shot ($N_\mrm{R}=1$), which even leaves a gain of $\SI{6}{\dB}$ when factoring in the loss of energy due to the four re-transmissions. We would like to argue, however, that an absolute power gain is not the key aim of the proposed technique. The aim is merely to wake up the EN device to enable CSI estimation in the initial-access phase. After the channel estimation, MRT will be used for WPT. 
Finally, we performed a Monte Carlo simulation ($N_\mrm{R}=10000$) to hint at the maximum achievable gain obtainable with beam diversity in the given simulation scenario.

Note that this paper only covers one part of 
the initial access to EN devices: the supply with sufficient power and first wake-up. 
To perform channel state estimation, the signal transmitted (e.g., backscattered) from the EN device must be 
successfully received by the \gls{rw} infrastructure.
Without the support of full-duplex communication, the signal may be received and processed by another array. 
After CSI estimation, the efficiency at which power can be transmitted using MRT depends on the quality of the CSI estimate~\cite{Mishra2019}. 
If perfect CSI was available,
the PG at the position $\bm{p}_\mrm{EN}$ yields $PG \approx \SI{-23.8}{\dB}$ in the simulated scenario.


That is, the EN device could receive $P_\mrm{RX} \approx \SI{10.9}{\dBm}$ when transmitting at $P_\mrm{TX} = \SI{3}{\watt}$. 
This is a significant amount of power transmitted over a distance of $y_\mrm{EN}=\SI{8.125}{\meter}$, which could open doors for a new generation of highly capable EN devices.
\changesBD{
Although this transmit power would result in a power density $S \approx \SI{9.99}{\watt\per\meter\squared}$ within the focal point that lies just below the 
human exposure limit of $\SI{10}{\watt\per\meter\squared}$, an exact evaluation will only be achievable with a realistic channel model and by accounting for any nonidealities.}
\changesBD{
For EN devices located at distances $y_\mrm{EN}<\SI{8.125}{\meter}$, the transmit power needs to be reduced accordingly for regulatory compliance.
Under the made model assumptions, the power density and the path gain are linearly dependent on each other through the relation
\begin{align}
    P_\mrm{RX} = S\, A = P_\mrm{TX}\, PG \,.
\end{align}
Therefore, Fig.~\ref{fig:MRT_full} indicates that the resulting power density levels close to the array will be much lower than in the focal point.
Using Nyquist-spaced arrays avoids 
spatial aliasing, which may result in grating lobes, or grating points, respectively, such that the only point of high power density is the focal point. 
}

\section{Conclusion and Outlook}\label{sec:conclusion}
There is great potential for indoor WPT using large or distributed arrays. 
We have shown that fading may severely impact the initial-access distance achievable with an indoor \gls{rw} deployment, but beam diversity can be exploited to overcome this impairment.
Environment-awareness helps focus multiple SMC beams to a point within the room simultaneously. 
By varying the phases of the beams, the necessary fading margin may be reduced at the price of a longer beam sweeping time and thus a higher amount of energy needed for initial access.
The lower necessary fading margin increases the initial access distance of EN devices. 

The parameters chosen for our simulations partially rely on assumptions made to demonstrate the concept of our beam diversity scheme. 
In the future, we will demonstrate the concept using real-life measurements. 
Furthermore, more elaborate schemes for exploiting beam diversity will be derived.

\vspace{0.4cm}

\bibliographystyle{IEEEtran}
\balance
\bibliography{IEEEabrv,ICC_2022_v2}


\end{document}

%% file: abbr.tex
\newacronym{aoa}{AOA}{angle-of-arrival}
\newacronym{aod}{AOD}{angle-of-departure}
\newacronym{asic}{ASIC}{application specific integrated circuit}
\newacronym{awgn}{AWGN}{additive white Gaussian noise}
\newacronym{ce}{CE}{channel estimation}
\newacronym{ced}{CED}{cumulative energy density}
\newacronym{cdf}{CDF}{cumulative distribution function}
\newacronym{csi}{CSI}{channel state information}
\newacronym{crlb}{CRLB}{Cram\'er-Rao lower bound}
\newacronym{dc}{DC}{direct current}
\newacronym{dm}{DM}{dense multipath}
\newacronym{eirp}{EIRP}{equivalent isotropically radiated power}
\newacronym{em}{EM}{electromagnetic}
\newacronym{en}{EN}{energy neutral}
\newacronym{e2e}{E2E}{End-to-End}
\newacronym{epu}{EPU}{edge processing unit}
\newacronym{fa}{FA}{federation anchor}
\newacronym{ic}{IC}{integrated circuit}
\newacronym{iot}{IoT}{Internet of things}
\newacronym{id}{ID}{information decoding}
\newacronym{ism}{ISM}{industrial, scientific and medical}
\newacronym{kpi}{KPI}{key performance indicator}
\newacronym{lca}{LCA}{life cycle assessment}
\newacronym{lis}{LIS}{large intelligent surface}
\newacronym{liion}{Li-ion}{lithium-ion}
\newacronym{los}{LoS}{line-of-sight}
\newacronym{lmo}{LMO}{lithium ion manganese oxide}
\newacronym{mf}{MF}{matched filter}
\newacronym{mmse}{MMSE}{minimum mean square error}
\newacronym{mmwave}{mmWave}{millimeter wave}
\newacronym{mimo}{MIMO}{multiple-input multiple-output}
\newacronym{miso}{MISO}{multiple-input single-output}
\newacronym{mpc}{MPC}{multipath component}
\newacronym{mrc}{MRC}{maximum ratio combining}
\newacronym{mrt}{MRT}{maximum ratio transmission}
\newacronym{ofdm}{OFDM}{orthogonal frequency-division multiplexing}
\newacronym{ota}{OTA}{over-the-air}
\newacronym{nb}{NB}{narrowband}
\newacronym{nr}{NR}{New Radio}
\newacronym{pa}{PA}{power amplifier}
\newacronym{pdp}{PDP}{power delay profile}
\newacronym{per}{PER}{packet error rate}
\newacronym{pl}{PL}{path loss}
\newacronym{pg}{PG}{path gain}
\newacronym{pc}{PC}{pilot count}
\newacronym{rcs}{RCS}{radar cross-section}
\newacronym{re}{RE}{radio element}
\newacronym{rf}{RF}{radio frequency}
\newacronym{rfid}{RFID}{radio frequency identification}
\newacronym{ris}{RIS}{reflective intelligent surface}
\newacronym{rms}{RMS}{root mean square}
\newacronym{rss}{RSS}{received signal strength}
\newacronym{rssi}{RSSI}{received signal strength indicator}
\newacronym{rw}{RW}{RadioWeaves}
\newacronym{sa}{SA}{synchronization anchor}
\newacronym{sdg}{SDG}{Sustainable Development Goal}
\newacronym{sdn}{SDN}{software-defined network}
\newacronym{sinr}{SINR}{signal-to-interference-plus-noise ratio}
\newacronym{siso}{SISO}{single-input single-output}
\newacronym{slam}{SLAM}{simultaneous localization and mapping}
\newacronym{smc}{SMC}{specular multipath component}
\newacronym{snr}{SNR}{signal-to-noise ratio}
\newacronym{svd}{SVD}{singular value decomposition}
\newacronym{tdd}{TDD}{time division duplexing}
\newacronym{tdoa}{TDOA}{time-difference-of-arrival}
\newacronym{toa}{TOA}{time-of-arrival}
\newacronym{ue}{UE}{user equipment}
\newacronym{uhf}{UHF}{ultra high frequency}
\newacronym{ula}{ULA}{uniform linear array}
\newacronym{upa}{UPA}{uniform planar array}
\newacronym{ura}{URA}{uniform rectangular array}
\newacronym{uwb}{UWB}{ultrawideband}
\newacronym{wb}{WB}{wideband}
\newacronym{wpt}{WPT}{wireless power transfer}
\newacronym{wsn}{WSN}{wireless sensor network}
\newacronym{zf}{ZF}{zero forcing}
\newacronym{csp}{CSP}{contact service point}
\newacronym{ecsp}{ECSP}{edge computing service point}
\newacronym{ap}{AP}{access point}

%% file: figures/setupICC.tex
%
%


\definecolor{mycolor1}{rgb}{0.52157,0.21961,0.35294}%
\definecolor{mycolor2}{rgb}{0.55294,0.75294,0.27059}%
\definecolor{mycolor3}{rgb}{1.00000,0.00000,1.00000}%
\definecolor{RDgreen}{rgb}{0.36471,0.42745,0.26667}
\begin{tikzpicture}

\begin{axis}[%
width=0.946\plotWidth,
height=0.667\plotWidth,
at={(0\plotWidth,0\plotWidth)},
plot box ratio=1.875 2.625 1,
xmin=-2,
xmax=13,
tick align=inside,		
axis line style = thick,	
xtick distance={3},				
ztick distance={2},				
xlabel style={font=\color{white!15!black}},
xlabel={$x$ in \SI{}{\meter}},
ymin=-2,
ymax=19,
ylabel style={font=\color{white!15!black}},
ylabel={$y$ in \SI{}{\meter}},
zmin=-3,
zmax=5,
zlabel style={font=\color{white!15!black}},
zlabel={$z$ in \SI{}{\meter}},
view={41.9872124958167}{24.0357477836919},
axis background/.style={fill=white},
xmajorgrids,
ymajorgrids,
zmajorgrids
]
\addplot3 [color=mycolor1, line width=1.0pt]
 table[row sep=crcr] {%
3	0	0\\
3	9	0\\
3	9	3.6\\
3	0	3.6\\
3	0	0\\
};
 
\addplot3[area legend, line width=1.5pt, draw=mycolor1, fill=mycolor1, fill opacity=0.2, forget plot, rounded corners=0.1pt, line cap=round]
table[row sep=crcr] {%
x	y	z\\
3	0	0\\
3	9	0\\
3	9	3.6\\
3	0	3.6\\
3	0	0\\
}--cycle;
\addplot3 [color=mycolor1, line width=1.5pt, rounded corners=0.1pt, line cap=round] 
 table[row sep=crcr] {%
8	0	0\\
8	9	0\\
8	9	3.6\\
8	0	3.6\\
8	0	0\\
};
 
\addplot3[area legend, line width=1.5pt, draw=mycolor1, fill=mycolor1, fill opacity=0.2, forget plot, rounded corners=0.1pt, line cap=round]
table[row sep=crcr] {%
x	y	z\\
8	0	0\\
8	9	0\\
8	9	3.6\\
8	0	3.6\\
8	0	0\\
}--cycle;

\addplot3 [color=mycolor1, line width=1.5pt, rounded corners=0.1pt, line cap=round] 
 table[row sep=crcr] {%
3	0	0\\
8	0	0\\
8	0	3.6\\
3	0	3.6\\
3	0	0\\
};
 
\addplot3[area legend, line width=1.5pt, draw=mycolor1, fill=mycolor1, fill opacity=0.2, forget plot, rounded corners=0.1pt, line cap=round]
table[row sep=crcr] {%
x	y	z\\
3	0	0\\
8	0	0\\
8	0	3.6\\
3	0	3.6\\
3	0	0\\
}--cycle;
\addplot3 [color=mycolor1, line width=1.5pt, rounded corners=0.1pt, line cap=round] 
 table[row sep=crcr] {%
3	9	0\\
8	9	0\\
8	9	3.6\\
3	9	3.6\\
3	9	0\\
};
 
\addplot3[area legend, line width=1.5pt, draw=mycolor1, fill=mycolor1, fill opacity=0.2, forget plot, rounded corners=0.1pt, line cap=round]
table[row sep=crcr] {%
x	y	z\\
3	9	0\\
8	9	0\\
8	9	3.6\\
3	9	3.6\\
3	9	0\\
}--cycle;
\addplot3 [color=mycolor1, line width=1.5pt, rounded corners=0.1pt, line cap=round] 
 table[row sep=crcr] {%
3	0	0\\
3	9	0\\
8	9	0\\
8	0	0\\
3	0	0\\
};
 
\addplot3[area legend, line width=1.5pt, draw=mycolor1, fill=mycolor1, fill opacity=0.2, forget plot, rounded corners=0.1pt, line cap=round]
table[row sep=crcr] {%
x	y	z\\
3	0	0\\
3	9	0\\
8	9	0\\
8	0	0\\
3	0	0\\
}--cycle;

 \addplot3 [color=red, line width=1.0pt, line cap=round]
 table[row sep=crcr] {%
5	0	1\\
6	0	1\\
};
 \node[right, align=left, font=\color{black!20!red}]
at (axis cs:6,0,1) {\footnotesize x};
\addplot3 [color=green, line width=1.0pt, line cap=round]
 table[row sep=crcr] {%
5	0	1\\
5	1	1\\
};
 \node[right, align=left, font=\color{black!20!green}]
at (axis cs:5,1,1) {\footnotesize y};
\addplot3 [color=blue, line width=1.0pt, line cap=round]
 table[row sep=crcr] {%
5	0	1\\
5	0	2\\
};
 \node[right, align=left, font=\color{black!20!blue}]
at (axis cs:5,0,2) {\footnotesize z};

\addplot3[area legend, draw=mycolor2, fill=mycolor2, fill opacity=0.3, forget plot]
table[row sep=crcr] {%
x	y	z\\
3.78125	-0.001	1.71875\\
3.78125	-0.001	1.71875\\
3.78125	-0.001	0.28125\\
6.21875	-0.001	0.28125\\
6.21875	-0.001	0.28125\\
6.21875	-0.001	1.71875\\
3.78125	-0.001	1.71875\\
}--cycle;

\addplot3 [color=mycolor2, line width=1.0pt, rounded corners=0.1pt, line cap=round] 
 table[row sep=crcr] {%
3.78125	-0.001	1.71875\\
3.78125	-0.001	1.71875\\
3.78125	-0.001	0.28125\\
6.21875	-0.001	0.28125\\
6.21875	-0.001	0.28125\\
6.21875	-0.001	1.71875\\
3.78125	-0.001	1.71875\\
};

\draw[thick][-{Latex[round]}](3,9,6) node[above, yshift=-2pt]{\small walls} --(3,8,4);		
\draw[thick][-{Latex[round]}](3,9,6)--(4,9,4);		

 \addplot3 [color=red, line width=1.0pt, line cap=round]
 table[row sep=crcr] {%
1	0	1\\
0	0	1\\
};
 \node[right, align=left, font=\color{black!20!red}]
at (axis cs:-1.5,0,1.1) {\footnotesize x};
\addplot3 [color=green, line width=1.0pt, line cap=round]
 table[row sep=crcr] {%
1	0	1\\
1	1	1\\
};
 \node[right, align=left, font=\color{black!20!green}]
at (axis cs:1,1,1) {\footnotesize y};
\addplot3 [color=blue, line width=1.0pt, line cap=round]
 table[row sep=crcr] {%
1	0	1\\
1	0	2\\
};
 \node[right, align=left, font=\color{black!20!blue}]
at (axis cs:1,0,2) {\footnotesize z};
\draw[thick][-{Latex[round]}](0.5,0,3.7) node[above, yshift=-4pt]{\small image array} --(0.5,0,1.8);		
\addplot3[area legend, draw=mycolor2, fill=mycolor2, fill opacity=0.3, forget plot]
table[row sep=crcr] {%
x	y	z\\
-0.21875	-0.001	1.71875\\
-0.21875	-0.001	1.71875\\
-0.21875	-0.001	0.28125\\
2.21875	-0.001	0.28125\\
2.21875	-0.001	0.28125\\
2.21875	-0.001	1.71875\\
-0.21875	-0.001	1.71875\\
}--cycle;
\addplot3 [color=mycolor2, line width=1.0pt, rounded corners=0.1pt, line cap=round] 
 table[row sep=crcr] {%
-0.21875	-0.001	1.71875\\
-0.21875	-0.001	1.71875\\
-0.21875	-0.001	0.28125\\
2.21875	-0.001	0.28125\\
2.21875	-0.001	0.28125\\
2.21875	-0.001	1.71875\\
-0.21875	-0.001	1.71875\\
};

 \addplot3 [color=red, line width=1.0pt, line cap=round]
 table[row sep=crcr] {%
11	0	1\\
10	0	1\\
};
 \node[right, align=left, font=\color{black!20!red}]
at (axis cs:8.5,0,1.1) {\footnotesize x};
\addplot3 [color=green, line width=1.0pt, line cap=round]
 table[row sep=crcr] {%
11	0	1\\
11	1	1\\
};
 \node[right, align=left, font=\color{black!20!green}]
at (axis cs:11,1,1) {\footnotesize y};
\addplot3 [color=blue, line width=1.0pt, line cap=round]
 table[row sep=crcr] {%
11	0	1\\
11	0	2\\
};
 \node[right, align=left, font=\color{black!20!blue}]
at (axis cs:11,0,2) {\footnotesize z};

\addplot3[area legend, draw=mycolor2, fill=mycolor2, fill opacity=0.3, forget plot]
table[row sep=crcr] {%
x	y	z\\
9.78125	-0.001	1.71875\\
9.78125	-0.001	1.71875\\
9.78125	-0.001	0.28125\\
12.21875	-0.001	0.28125\\
12.21875	-0.001	0.28125\\
12.21875	-0.001	1.71875\\
9.78125	-0.001	1.71875\\
}--cycle;
\addplot3 [color=mycolor2, line width=1.0pt, rounded corners=0.1pt, line cap=round] 
 table[row sep=crcr] {%
9.78125	-0.001	1.71875\\
9.78125	-0.001	1.71875\\
9.78125	-0.001	0.28125\\
12.21875	-0.001	0.28125\\
12.21875	-0.001	0.28125\\
12.21875	-0.001	1.71875\\
9.78125	-0.001	1.71875\\
};

\addplot3[area legend, draw=mycolor2, fill=mycolor2, fill opacity=0.3, forget plot]
table[row sep=crcr] {%
x	y	z\\
3.78125	17.999	1.71875\\
3.78125	17.999	1.71875\\
3.78125	17.999	0.28125\\
6.21875	17.999	0.28125\\
6.21875	17.999	0.28125\\
6.21875	17.999	1.71875\\
3.78125	17.999	1.71875\\
}--cycle;
\addplot3 [color=mycolor2, line width=1.0pt, rounded corners=0.1pt, line cap=round] 
 table[row sep=crcr] {%
3.78125	17.999	1.71875\\
3.78125	17.999	1.71875\\
3.78125	17.999	0.28125\\
6.21875	17.999	0.28125\\
6.21875	17.999	0.28125\\
6.21875	17.999	1.71875\\
3.78125	17.999	1.71875\\
};

  \addplot3 [color=red, line width=1.0pt, line cap=round]
 table[row sep=crcr] {%
5	18	1\\
6	18	1\\
};
 \node[right, align=left, font=\color{black!20!red}]
at (axis cs:6,18,1) {\footnotesize x};
\addplot3 [color=green, line width=1.0pt, line cap=round]
 table[row sep=crcr] {%
5	18	1\\
5	17	1\\
};
 \node[right, align=left, font=\color{black!20!green}]
at (axis cs:5.5,16,1) {\footnotesize y};
\addplot3 [color=blue, line width=1.0pt, line cap=round]
 table[row sep=crcr] {%
5	18	1\\
5	18	2\\
};
 \node[right, align=left, font=\color{black!20!blue}]
at (axis cs:5,18,2) {\footnotesize z};

 \addplot3 [color=red, line width=1.0pt, line cap=round]
 table[row sep=crcr] {%
5	0	-1\\
6	0	-1\\
};
 \node[right, align=left, font=\color{black!20!red}]
at (axis cs:6,0,-1) {\footnotesize x};
\addplot3 [color=green, line width=1.0pt, line cap=round]
 table[row sep=crcr] {%
5	0	-1\\
5	1	-1\\
};
 \node[right, align=left, font=\color{black!20!green}]
at (axis cs:5,1,-1) {\footnotesize y};
\addplot3 [color=blue, line width=1.0pt, line cap=round]
 table[row sep=crcr] {%
5	0	-1\\
5	0	-2\\
};
 \node[right, align=left, font=\color{black!20!blue}]
at (axis cs:5,0,-2) {\footnotesize z};

\addplot3[area legend, draw=mycolor2, fill=mycolor2, fill opacity=0.3, forget plot]
table[row sep=crcr] {%
x	y	z\\
3.78125	-0.001	-0.28125\\
3.78125	-0.001	-0.28125\\
3.78125	-0.001	-1.71875\\
6.21875	-0.001	-1.71875\\
6.21875	-0.001	-1.71875\\
6.21875	-0.001	-0.28125\\
3.78125	-0.001	-0.28125\\
}--cycle;
\addplot3 [color=mycolor2, line width=1.0pt, rounded corners=0.1pt, line cap=round] 
 table[row sep=crcr] {%
3.78125	-0.001	-0.28125\\
3.78125	-0.001	-0.28125\\
3.78125	-0.001	-1.71875\\
6.21875	-0.001	-1.71875\\
6.21875	-0.001	-1.71875\\
6.21875	-0.001	-0.28125\\
3.78125	-0.001	-0.28125\\
};

\draw[thick][-{Latex[round]}](0.5,1,-2.5) node[below, yshift=2pt]{\small RW array} --(3.7,0,0.25);		
\draw[thick][-{Latex[round]}](11,8.1,4) node[right]{\small scatter points} --(6.5,8.1,1.5);		

 \addplot3[scatter, only marks, mark=o, color=black, mark options={}, scatter/use mapped color=black, visualization depends on={\thisrow{size} \as \perpointmarksize}, scatter/@pre marker code/.append style={/tikz/mark size=\perpointmarksize}, forget plot] table[row sep=crcr]{%
x	y	z	size\\
5.33051849524216	8.65076578477899	1.20768311101162	0.862682538024909\\
4.67395112920194	8.64311551419747	0.479153685022454	0.946812334471243\\
5.00058160996353	8.71726030114687	0.66216551596227	0.962631561479956\\
4.89325658972314	8.79969296991617	0.45624761740377	1.00364834417035\\
4.96990957833925	8.82968707310526	0.314709534615471	0.908507615632375\\
5.34083140011076	8.66288358608921	0.746979176882175	0.936208000504268\\
4.69087865413012	8.80000524662204	1.35654306995795	0.973369288394909\\
5.36960869241959	8.72233085324363	1.02950870504394	1.0154881952361\\
4.89636432595591	8.69287146806877	1.0212777839069	0.873626493585031\\
4.65303970151706	8.63905181065907	1.45089779845778	0.859157835285881\\
5.22143247563378	8.8148639129701	0.664053598037616	0.915143061728701\\
5.41090983916627	8.70500930561373	0.787103081067935	0.934009462308485\\
4.44549987964596	8.81016883989523	1.10152392413244	0.957370716954673\\
4.9145637121342	8.73471685281564	1.26519231036268	0.946171832028476\\
4.48154470048046	8.86253324756423	1.07608571815938	0.910125177292902\\
5.00871411250911	8.78555638677071	0.61917430148782	0.972944534461818\\
5.41265906986815	8.65445734723264	1.37807179341522	0.943344144053348\\
4.49630881827245	8.66619953226826	0.644401110757165	0.919653768109294\\
5.27679180032121	8.75166414170255	1.21546943835024	0.941696745784272\\
4.59625965603702	8.62808927039878	0.874627484518682	0.997618440607122\\
4.78880632000664	8.8847036135919	1.48289081820724	0.960230892236832\\
4.52100384758846	8.72259052787842	1.27631099108121	0.918557629324985\\
5.20694906451148	8.6899052440143	1.03106897315537	0.857344019526891\\
4.73897288760911	8.85892759362861	0.557062163291679	0.924494855173765\\
5.2606622771435	8.69373116090866	0.424158333733894	0.945104772471118\\
4.71956796161488	8.65842939920695	0.544759076646777	0.916868883282871\\
5.30215303566958	8.76004523342578	0.329685391566916	0.958586570828934\\
5.35131637889882	8.69438773751231	0.507158799790594	0.886196047316502\\
4.7800249042469	8.56170878341988	0.602263040234301	0.920285146040862\\
5.35015589386988	8.74803648759227	1.16488810422527	0.961807558643678\\
5.18095431977749	8.8636455403235	1.04643173369484	0.863583502761723\\
5.10127594875469	8.78934361027084	0.319161626566174	0.958699816316298\\
5.25989512742002	8.71094682279858	0.527859533842754	0.942006394788954\\
4.95196031067232	8.7038749471006	1.38471292942231	1.00883987189335\\
5.66318778738451	8.84731621274286	1.09962017387823	0.909980819403081\\
5.01456892766443	8.60857686236445	0.687021606980144	0.860364641180672\\
5.0144205058029	8.55267204912512	0.662337721469017	0.971800830344783\\
5.52099267972792	8.65878106057797	1.32871244878068	0.98862389816905\\
};
\addplot3 [color=mycolor2]
 table[row sep=crcr] {%
5	8.125	0\\
5	8.125	1\\
};


\draw[thick][-{Latex[round]}](7,7.8,-2) node[below, xshift=2pt, xshift=15pt]{\small EN device} --(5.2,7.9,0.75);		
 \addplot3 [color=mycolor2, line width=1.0pt, only marks, mark=o, mark options={solid}]
 table[row sep=crcr] {%
5	8.125	1\\
};
\end{axis}
\end{tikzpicture}%

%% file: figures/MRT_full.tex
%
%
\definecolor{RDgreen}{rgb}{0.36471,0.42745,0.26667}
\definecolor{ArrayGreen}{rgb}{0.55294,0.75294,0.27059}

\begin{tikzpicture}

\path[use as bounding box] (-0.75cm,-0.15cm) rectangle (8.25cm,4.5cm);

\begin{axis}[%
axis line style = thick,	    
xtick distance={1},				
ytick distance={2},				
unit vector ratio={1 1},
rotate=-90,
width=0.9\plotWidth,
height=0.5\plotWidth,
at={(0\plotWidth,0\plotWidth)},
scale only axis,
point meta min=-48.8751231246868,		
point meta max=-23.8751231246868,		
axis on top,
xmin=3,
xmax=8,
xlabel style={font=\color{white!15!black},rotate=90,anchor=base,yshift=0.25cm},		
xticklabel style = {anchor=base,xshift=-0.2cm,yshift=-1mm},							
xlabel={$x$ in \SI{}{m}},													
ymin=0,
ymax=9,
ylabel style={font=\color{white!15!black},rotate=-90,anchor=base,yshift=0.4cm},		
yticklabel style = {anchor=base,yshift=0.15cm},									
ylabel={$y$ in \SI{}{m}},
axis background/.style={fill=white},
xmajorgrids,
ymajorgrids,
colormap={mymap}{[1pt] rgb(0pt)=(1,1,1); rgb(26pt)=(0.51634,0.620915,1); rgb(27pt)=(0.497738,0.606335,1); rgb(34pt)=(0.367521,0.504274,1); rgb(35pt)=(0.348919,0.489693,1); rgb(39pt)=(0.27451,0.431373,1); rgb(40pt)=(0.317186,0.405998,0.941176); rgb(41pt)=(0.359862,0.380623,0.882353); rgb(43pt)=(0.445213,0.329873,0.764706); rgb(44pt)=(0.487889,0.304498,0.705882); rgb(45pt)=(0.530565,0.279123,0.647059); rgb(46pt)=(0.573241,0.253749,0.588235); rgb(47pt)=(0.615917,0.228374,0.529412); rgb(48pt)=(0.658593,0.202999,0.470588); rgb(49pt)=(0.701269,0.177624,0.411765); rgb(50pt)=(0.743945,0.152249,0.352941); rgb(51pt)=(0.786621,0.126874,0.294118); rgb(52pt)=(0.829296,0.101499,0.235294); rgb(53pt)=(0.871972,0.0761246,0.176471); rgb(54pt)=(0.914648,0.0507497,0.117647); rgb(55pt)=(0.957324,0.0253749,0.0588235); rgb(56pt)=(1,0,0); rgb(58pt)=(1,0.285714,0); rgb(59pt)=(1,0.428571,0); rgb(60pt)=(1,0.571429,0); rgb(63pt)=(1,1,0)},
colorbar,
colorbar style={ylabel style={font=\color{white!15!black}},width=0.3cm,ytick pos=right, ytick distance={5}, tick align=center, axis line style = thick,	
ylabel={$PG$ in \SI{}{\dB}}}
]
\addplot[scatter, only marks, mark=o, color=black, mark options={}, scatter/use mapped color=black, visualization depends on={\thisrow{size} \as \perpointmarksize}, scatter/@pre marker code/.append style={/tikz/mark size=\perpointmarksize}, forget plot] table[row sep=crcr]{%
x	y	size\\
5.33051849524216	8.65076578477899	1.6139312454209\\
4.67395112920194	8.64311551419747	1.7713236825815\\
5.00058160996353	8.71726030114687	1.80091874637661\\
4.89325658972314	8.79969296991617	1.87765412034421\\
4.96990957833925	8.82968707310526	1.69966211548563\\
5.34083140011076	8.66288358608921	1.75148479032183\\
4.69087865413012	8.80000524662204	1.82100719399085\\
5.36960869241959	8.72233085324363	1.89980445344344\\
4.89636432595591	8.69287146806877	1.63440551150193\\
4.65303970151706	8.63905181065907	1.60733713040106\\
5.22143247563378	8.8148639129701	1.71207589843606\\
5.41090983916627	8.70500930561373	1.74737170198165\\
4.44549987964596	8.81016883989523	1.79107660748726\\
4.9145637121342	8.73471685281564	1.77012541223339\\
4.48154470048046	8.86253324756423	1.70268829625346\\
5.00871411250911	8.78555638677071	1.8202125521452\\
5.41265906986815	8.65445734723264	1.76483529243357\\
4.49630881827245	8.66619953226826	1.72051465736031\\
5.27679180032121	8.75166414170255	1.76175329248235\\
4.59625965603702	8.62808927039878	1.86637320373977\\
4.78880632000664	8.8847036135919	1.79642750547324\\
4.52100384758846	8.72259052787842	1.7184639694707\\
5.20694906451148	8.6899052440143	1.60394379183463\\
4.73897288760911	8.85892759362861	1.72957150194771\\
5.2606622771435	8.69373116090866	1.76812912659593\\
4.71956796161488	8.65842939920695	1.71530461491927\\
5.30215303566958	8.76004523342578	1.79335126180219\\
5.35131637889882	8.69438773751231	1.65792099328583\\
4.7800249042469	8.56170878341988	1.72169585731106\\
5.35015589386988	8.74803648759227	1.79937717822706\\
5.18095431977749	8.8636455403235	1.61561679610226\\
5.10127594875469	8.78934361027084	1.79356312470935\\
5.25989512742002	8.71094682279858	1.76233259272519\\
4.95196031067232	8.7038749471006	1.88736657937091\\
5.66318778738451	8.84731621274286	1.70241822737107\\
5.01456892766443	8.60857686236445	1.60959485749639\\
5.0144205058029	8.55267204912512	1.8180728776663\\
5.52099267972792	8.65878106057797	1.84954595566274\\
};
\addplot [forget plot] graphics [xmin=2.9937343358396, xmax=8.0062656641604, ymin=0.0512568332137636, ymax=9.01119992886957] {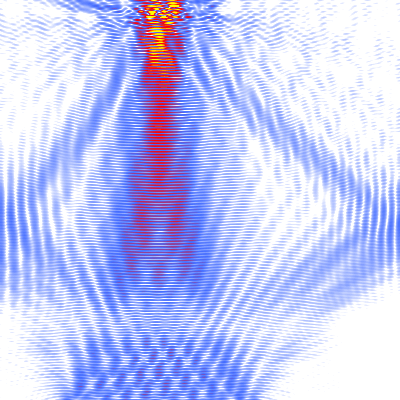};		
\addplot [color=ArrayGreen, line width=1.5pt, only marks, mark=o, mark options={solid, ArrayGreen}, forget plot]	
  table[row sep=crcr]{%
5	8.125\\
};
\addplot[color=ArrayGreen, line width=1.0pt, line cap=round]
 table[row sep=crcr] {%
3.75	0.05\\
6.25	0.05\\
};
\draw[thick][-{Latex[round]}](5.75,7) node[left, xshift=2pt]{\small EN device} --(5.1,7.95);		
\end{axis}
\end{tikzpicture}%

%% file: figures/CB.tex
%
%
\begin{tikzpicture}

\begin{axis}[%
width=0.485\plotWidth,
height=4\plotWidth,
at={(0\plotWidth,0\plotWidth)},
scale only axis,
xmin=0,
xmax=1.02,
ymin=-1,
ymax=1,
zmin=-62.4038938574353,
zmax=-22.4038938574353,
zlabel style={font=\color{white!15!black}},
zlabel={$PG$ in \SI{}{\dB}},
axis line style = thick,	
view={0}{0},
axis background/.style={fill=white},
axis z line*=right,			
xticklabel=\empty,			
tick align=center,		    
xtick=\empty,               
y axis line style= { draw opacity=0 }, 
z axis line style= { draw opacity=0 }, 
ztick distance={5},				
yticklabel pos=right,
xmajorgrids,
ymajorgrids,
zmajorgrids
]

\addplot3[%
surf,
shader=flat, z buffer=sort, colormap={mymap}{[1pt] rgb(0pt)=(1,1,1); rgb(26pt)=(0.51634,0.620915,1); rgb(27pt)=(0.497738,0.606335,1); rgb(34pt)=(0.367521,0.504274,1); rgb(35pt)=(0.348919,0.489693,1); rgb(39pt)=(0.27451,0.431373,1); rgb(40pt)=(0.317186,0.405998,0.941176); rgb(41pt)=(0.359862,0.380623,0.882353); rgb(43pt)=(0.445213,0.329873,0.764706); rgb(44pt)=(0.487889,0.304498,0.705882); rgb(45pt)=(0.530565,0.279123,0.647059); rgb(46pt)=(0.573241,0.253749,0.588235); rgb(47pt)=(0.615917,0.228374,0.529412); rgb(48pt)=(0.658593,0.202999,0.470588); rgb(49pt)=(0.701269,0.177624,0.411765); rgb(50pt)=(0.743945,0.152249,0.352941); rgb(51pt)=(0.786621,0.126874,0.294118); rgb(52pt)=(0.829296,0.101499,0.235294); rgb(53pt)=(0.871972,0.0761246,0.176471); rgb(54pt)=(0.914648,0.0507497,0.117647); rgb(55pt)=(0.957324,0.0253749,0.0588235); rgb(56pt)=(1,0,0); rgb(58pt)=(1,0.285714,0); rgb(59pt)=(1,0.428571,0); rgb(60pt)=(1,0.571429,0); rgb(63pt)=(1,1,0)}, mesh/rows=2]
table[row sep=crcr, point meta=\thisrow{c}] {%
x	y	z	c\\
0	-0.0001	-22.4038938574353	-22.4038938574353\\
0	-0.0001	-23.038814492356	-23.038814492356\\
0	-0.0001	-23.6737351272766	-23.6737351272766\\
0	-0.0001	-24.3086557621972	-24.3086557621972\\
0	-0.0001	-24.9435763971179	-24.9435763971179\\
0	-0.0001	-25.5784970320385	-25.5784970320385\\
0	-0.0001	-26.2134176669591	-26.2134176669591\\
0	-0.0001	-26.8483383018798	-26.8483383018798\\
0	-0.0001	-27.4832589368004	-27.4832589368004\\
0	-0.0001	-28.1181795717211	-28.1181795717211\\
0	-0.0001	-28.7531002066417	-28.7531002066417\\
0	-0.0001	-29.3880208415623	-29.3880208415623\\
0	-0.0001	-30.022941476483	-30.022941476483\\
0	-0.0001	-30.6578621114036	-30.6578621114036\\
0	-0.0001	-31.2927827463242	-31.2927827463242\\
0	-0.0001	-31.9277033812449	-31.9277033812449\\
0	-0.0001	-32.5626240161655	-32.5626240161655\\
0	-0.0001	-33.1975446510861	-33.1975446510861\\
0	-0.0001	-33.8324652860068	-33.8324652860068\\
0	-0.0001	-34.4673859209274	-34.4673859209274\\
0	-0.0001	-35.102306555848	-35.102306555848\\
0	-0.0001	-35.7372271907687	-35.7372271907687\\
0	-0.0001	-36.3721478256893	-36.3721478256893\\
0	-0.0001	-37.0070684606099	-37.0070684606099\\
0	-0.0001	-37.6419890955306	-37.6419890955306\\
0	-0.0001	-38.2769097304512	-38.2769097304512\\
0	-0.0001	-38.9118303653718	-38.9118303653718\\
0	-0.0001	-39.5467510002925	-39.5467510002925\\
0	-0.0001	-40.1816716352131	-40.1816716352131\\
0	-0.0001	-40.8165922701337	-40.8165922701337\\
0	-0.0001	-41.4515129050544	-41.4515129050544\\
0	-0.0001	-42.086433539975	-42.086433539975\\
0	-0.0001	-42.7213541748957	-42.7213541748957\\
0	-0.0001	-43.3562748098163	-43.3562748098163\\
0	-0.0001	-43.9911954447369	-43.9911954447369\\
0	-0.0001	-44.6261160796576	-44.6261160796576\\
0	-0.0001	-45.2610367145782	-45.2610367145782\\
0	-0.0001	-45.8959573494988	-45.8959573494988\\
0	-0.0001	-46.5308779844195	-46.5308779844195\\
0	-0.0001	-47.1657986193401	-47.1657986193401\\
0	-0.0001	-47.8007192542607	-47.8007192542607\\
0	-0.0001	-48.4356398891814	-48.4356398891814\\
0	-0.0001	-49.070560524102	-49.070560524102\\
0	-0.0001	-49.7054811590226	-49.7054811590226\\
0	-0.0001	-50.3404017939433	-50.3404017939433\\
0	-0.0001	-50.9753224288639	-50.9753224288639\\
0	-0.0001	-51.6102430637845	-51.6102430637845\\
0	-0.0001	-52.2451636987052	-52.2451636987052\\
0	-0.0001	-52.8800843336258	-52.8800843336258\\
0	-0.0001	-53.5150049685464	-53.5150049685464\\
0	-0.0001	-54.1499256034671	-54.1499256034671\\
0	-0.0001	-54.7848462383877	-54.7848462383877\\
0	-0.0001	-55.4197668733083	-55.4197668733083\\
0	-0.0001	-56.054687508229	-56.054687508229\\
0	-0.0001	-56.6896081431496	-56.6896081431496\\
0	-0.0001	-57.3245287780703	-57.3245287780703\\
0	-0.0001	-57.9594494129909	-57.9594494129909\\
0	-0.0001	-58.5943700479115	-58.5943700479115\\
0	-0.0001	-59.2292906828322	-59.2292906828322\\
0	-0.0001	-59.8642113177528	-59.8642113177528\\
0	-0.0001	-60.4991319526734	-60.4991319526734\\
0	-0.0001	-61.1340525875941	-61.1340525875941\\
0	-0.0001	-61.7689732225147	-61.7689732225147\\
0	-0.0001	-62.4038938574353	-62.4038938574353\\
1	-0.0001	-22.4038938574353	-22.4038938574353\\
1	-0.0001	-23.038814492356	-23.038814492356\\
1	-0.0001	-23.6737351272766	-23.6737351272766\\
1	-0.0001	-24.3086557621972	-24.3086557621972\\
1	-0.0001	-24.9435763971179	-24.9435763971179\\
1	-0.0001	-25.5784970320385	-25.5784970320385\\
1	-0.0001	-26.2134176669591	-26.2134176669591\\
1	-0.0001	-26.8483383018798	-26.8483383018798\\
1	-0.0001	-27.4832589368004	-27.4832589368004\\
1	-0.0001	-28.1181795717211	-28.1181795717211\\
1	-0.0001	-28.7531002066417	-28.7531002066417\\
1	-0.0001	-29.3880208415623	-29.3880208415623\\
1	-0.0001	-30.022941476483	-30.022941476483\\
1	-0.0001	-30.6578621114036	-30.6578621114036\\
1	-0.0001	-31.2927827463242	-31.2927827463242\\
1	-0.0001	-31.9277033812449	-31.9277033812449\\
1	-0.0001	-32.5626240161655	-32.5626240161655\\
1	-0.0001	-33.1975446510861	-33.1975446510861\\
1	-0.0001	-33.8324652860068	-33.8324652860068\\
1	-0.0001	-34.4673859209274	-34.4673859209274\\
1	-0.0001	-35.102306555848	-35.102306555848\\
1	-0.0001	-35.7372271907687	-35.7372271907687\\
1	-0.0001	-36.3721478256893	-36.3721478256893\\
1	-0.0001	-37.0070684606099	-37.0070684606099\\
1	-0.0001	-37.6419890955306	-37.6419890955306\\
1	-0.0001	-38.2769097304512	-38.2769097304512\\
1	-0.0001	-38.9118303653718	-38.9118303653718\\
1	-0.0001	-39.5467510002925	-39.5467510002925\\
1	-0.0001	-40.1816716352131	-40.1816716352131\\
1	-0.0001	-40.8165922701337	-40.8165922701337\\
1	-0.0001	-41.4515129050544	-41.4515129050544\\
1	-0.0001	-42.086433539975	-42.086433539975\\
1	-0.0001	-42.7213541748957	-42.7213541748957\\
1	-0.0001	-43.3562748098163	-43.3562748098163\\
1	-0.0001	-43.9911954447369	-43.9911954447369\\
1	-0.0001	-44.6261160796576	-44.6261160796576\\
1	-0.0001	-45.2610367145782	-45.2610367145782\\
1	-0.0001	-45.8959573494988	-45.8959573494988\\
1	-0.0001	-46.5308779844195	-46.5308779844195\\
1	-0.0001	-47.1657986193401	-47.1657986193401\\
1	-0.0001	-47.8007192542607	-47.8007192542607\\
1	-0.0001	-48.4356398891814	-48.4356398891814\\
1	-0.0001	-49.070560524102	-49.070560524102\\
1	-0.0001	-49.7054811590226	-49.7054811590226\\
1	-0.0001	-50.3404017939433	-50.3404017939433\\
1	-0.0001	-50.9753224288639	-50.9753224288639\\
1	-0.0001	-51.6102430637845	-51.6102430637845\\
1	-0.0001	-52.2451636987052	-52.2451636987052\\
1	-0.0001	-52.8800843336258	-52.8800843336258\\
1	-0.0001	-53.5150049685464	-53.5150049685464\\
1	-0.0001	-54.1499256034671	-54.1499256034671\\
1	-0.0001	-54.7848462383877	-54.7848462383877\\
1	-0.0001	-55.4197668733083	-55.4197668733083\\
1	-0.0001	-56.054687508229	-56.054687508229\\
1	-0.0001	-56.6896081431496	-56.6896081431496\\
1	-0.0001	-57.3245287780703	-57.3245287780703\\
1	-0.0001	-57.9594494129909	-57.9594494129909\\
1	-0.0001	-58.5943700479115	-58.5943700479115\\
1	-0.0001	-59.2292906828322	-59.2292906828322\\
1	-0.0001	-59.8642113177528	-59.8642113177528\\
1	-0.0001	-60.4991319526734	-60.4991319526734\\
1	-0.0001	-61.1340525875941	-61.1340525875941\\
1	-0.0001	-61.7689732225147	-61.7689732225147\\
1	-0.0001	-62.4038938574353	-62.4038938574353\\
};
\draw [draw=black, thick] (0.015,0.98,-62.4038938574353) rectangle (1,1,-22.4038938574353);    
\end{axis}
\end{tikzpicture}%